# Comparing Photorealistic and Animated Embodied Conversational Agents in Serious Games: An Empirical Study on User Experience

Danai Korre

The University of Edinburgh, Edinburgh, UK
d.korre@ed.ac.uk

**Abstract.** Embodied conversational agents (ECAs) are paradigms of conversational user interfaces in the form of embodied characters. While ECAs offer various manipulable features, this paper focuses on a study conducted to explore two distinct levels of presentation realism. The two agent versions are photorealistic and animated. The study aims to provide insights and design suggestions for speech-enabled ECAs within serious game environments. A within-subjects 2 x 2 factorial design was employed for this research with a cohort of 36 participants balanced for gender. The results showed that both the photorealistic and the animated versions were perceived as highly usable with overall mean scores of 5.76 and 5.71 respectively. However, 69.4% of participants stated they preferred the photorealistic version; 25% of participants stated they preferred the animated version; and 5.6% of participants had no stated preference. The photorealistic agents were perceived as more realistic and human-like while the animated characters made the task feel more like a game. Even though the agents' realism had no significant effect on usability, it positively influenced participants' perceptions of the agent. This research aims to lay the groundwork for future studies on the impact of ECA realism in serious games across diverse contexts.

## 1 Introduction

There are two information processors in human-computer interaction (HCI): a computer and a human. These two entities communicate with each other through a constrained interface. Understanding and studying the design of this interface is crucial for overcoming its inherent limitations [1].

Conversational user interfaces (CUIs) are one of the modes of interaction that can be employed in HCI. Recent technological developments in artificial intelligence (AI) and the increasing popularity of voice assistants such as Google Home and Amazon Echo, have sparked a growing interest in conversational user interfaces (CUIs) [2]. Em-bodied conversational agents (ECAs) are CUI paradigms in the form of virtual characters designed to emulate human-to-human interaction using a combination of verbal and non-verbal communication [3]. In this paper's study, the focus lies on the utilization of speech-enabled ECAs.

Previous work has shown that the interaction with spoken dialogue systems, either in the form of an embodied agent or not, is still inferior to other approaches that allow direct manipulation, despite the theoretical advances of ECAs and dialogue systems [4]. The reasons for the reluctance of using ECAs in multimodal HCI are multi-fold. On the human side of the interaction, when communicating audio-visually via speech people convey extra-linguistic information instinctively. In HCI this information that works complementary to speech is not reliably communicated since machines are usually not able to interpret and extract this type of information. On the machine side of the interaction, current technology is still somewhat limited to generating extra-linguistic information the way humans do. Furthermore, such attempts are not always interpreted correctly by users [3,4].

Considering the factors mentioned earlier, the evaluation of these ECAs still poses several challenges, especially within the encompassing design process that entails prototyping and testing. Designing ECAs is a time-consuming task that involves multidisciplinary fields such as human-computer interaction (HCI), computer science, behavioral science, psychology, and linguistics [3]. This paper aims to shed light on these interaction paradigms by evaluating the design of speech-enabled ECAs in the context of a serious game application and providing insights based on both quantitative and qualitative data and observations.

## 1.1 Embodied conversational agents, their roles, and characteristics

Embodied conversational agents, as they exist today, are the outcome of various interdisciplinary fields. The specific disciplines involved in the development of each ECA vary depending on its functionalities and modes of interaction. For instance, an ECA utilizing speech input requires speech recognition and speech-to-text technology, whereas ECAs with text input do not. Overall, ECAs inherently embody a multidisciplinary nature, as depicted in Fig.1

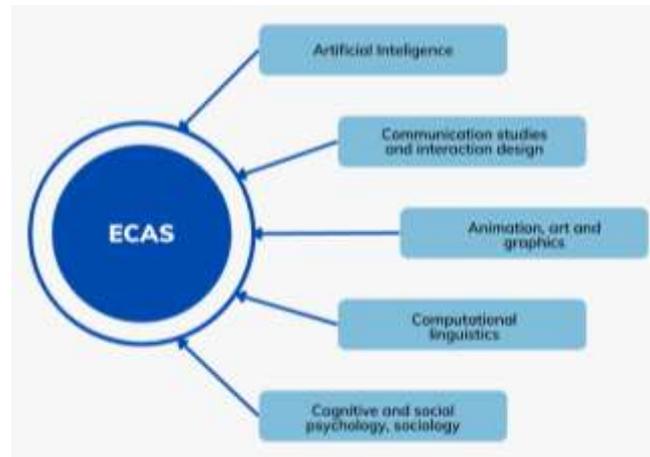

**Fig. 1.** Contributing disciplines to the field of ECAs

There are many theoretical advantages regarding ECAs and spoken dialogue systems since they provide a more "natural interaction" [3, 8]. Their visual design has been deemed important beyond the aesthetic aspect of it [4].

Previous research has identified the possibilities of using ECAs for mission rehearsal training [5], for military leadership and cultural training [6, 7], as museum guides and in installations [8], as sales agents [9-11], as medical advisers [12, 13], as companions [14], e-commerce and finance [15-17], as TV-style presenters [18], for psychological support [19] and in various other roles [3].

Embodied conversational agents serve as animated pedagogical agents in a variety of roles [20-23] when used in an educational context. According to Kim and Baylor, [24], pedagogical agents have been found in four major roles: 1) an expert who provides information, 2) a mentor who advises, 3) a motivator who encourages, and 4) a companion who collaborates.

According to De Vos, [25], ECAs share the following five features: anthropomorphic appearance, virtual body that is used for communication purposes, natural communication protocols, multimodality, and a social role.

### 1.2 Embodied conversational agent appearance

According to Nass [26], appearance influences people's cognitive assessments. According to Churchill et al. [27, 28], the character's appearance plays a significant role in naming which design aspects should be examined for situated conversational characters. Gulz and Haake [29] proposed five dimensions of agent look to be explored:
1. the degree or level of human likeness,
2. the degree of stability versus changeability in terms of appearance,
3. animation or movement versus static or immobility,
4. 2D or 3D visual rendering and
5. the degree of realism ranging from photorealism to stylized.

The presented study focuses on the fifth dimension of the degree of realism.

Although conducting a comprehensive review of the literature on ECA realism is beyond the scope of this paper, indicative previous research examining the level of ECA realism has provided mixed findings. In a study comparing two levels of pedagogical agent realism, specifically a cartoon-like animal and an anthropomorphic agent with a voice-only tutor as a control, no definitive advantage was found for the agents in terms of the learning effect. However, the results did show that the appearance of the agent does have an impact [30]. In another study looking at the design implications of a virtual learning companion robot, the participants preferred the robot-like appearance over the human-like agent [31]. Another study with three degrees of agent representation, computer vs. virtual vs. human agent, showed that participants' decision to trust the agent by adopting task-specific advice was not affected [32]. Finally, a research study on the effect of embodiment for a group facilitation agent, showed that the agent embodiment, non-verbal communication, and human-like face had an effect improving user perceptions [33].

Previous research has focused on the design aspects of ECAs but there are limited empirical evaluations regarding their effectiveness specifically focusing on serious games [3]. Given the increasing importance of usability in modern development processes, it is crucial to evaluate the introduction of ECAs in such environments. Failing to do so can lead to potential issues. Therefore, it is highly necessary to investigate the specific aspects of ECAs that can influence usability within serious game environments, surpassing current interaction paradigms. The study is a comparative evaluation between two design styles commonly used in gaming, photorealism, and animated design, and investigates how the different styles affect usability through an empirical study.

## 2  Research question/hypothesis

The primary objective of this study is to examine how ECA design decisions affect the usability of a serious game application. Specifically, this experiment investigates user reactions to two versions of multimodal interaction with ECAs. One version contains photorealistic ECAs and the second contains animated ECAs. This experiment aims to present an examination of how and if ECAs' aesthetics affect the usability of the application.

## 3  Methodology

A user study was conducted to compare two ECA designs. The study aimed to compare the two ECA designs in terms of user preference and usability.

### 3.1  Participants

A cohort of 36 participants (18 females and 18 males; aged 18 to 40) was recruited to take part in the experiments. The participants were balanced for gender, version order, and shopping list order. Participants were selected from a list of bank customers who signed up for user studies. All participants were comfortable using a desktop computer. There were no prerequisites for taking part in the study. The only limitation was that participants had to be between 18 and 40 years of age. The age limit was calculated based on the context of the game (on pre-decimalized currency) that was used up until the 70s therefore, it is highly unlikely for someone under 40 years old to have used the old money system.

### 3.2  Apparatus

The experiment was conducted on a desktop computer. The interface was presented on a Dell Precision T3600 with NVIDIA Quadro K2000 3D vision Pro graphics card, a standard 24" PC monitor with a resolution of 2220 × 1080 pixels, and an Intel® Xeon® Processor E5-1603, 10M Cache, 2.80 GHz, 0.0 GT/s Intel® QPI processor.
The software was developed using Unity. The overall application was not specifically designed for the evaluations reported in this paper, but the different ECA designs were designed specifically for the study.

The game that was used is called Moneyworld. Moneyworld is a computer-based game where the player is asked to travel back in time and use the pre-decimal money system to purchase a list of items in a corner store back in the 60s. All the transactions need to be made by using the following pre-decimal coins: penny, threepence, sixpence, shilling, and florin. The scene is composed of the corner store, the shopkeeper with whom the player interacts to buy the items that Alex dictates, Alex on the top right corner, and on the left side is the inventory of the items purchased and the rewards system. When asked to pay for the items, the screen fades to dark and the user is presented with the coins (Fig.2, Fig.3).

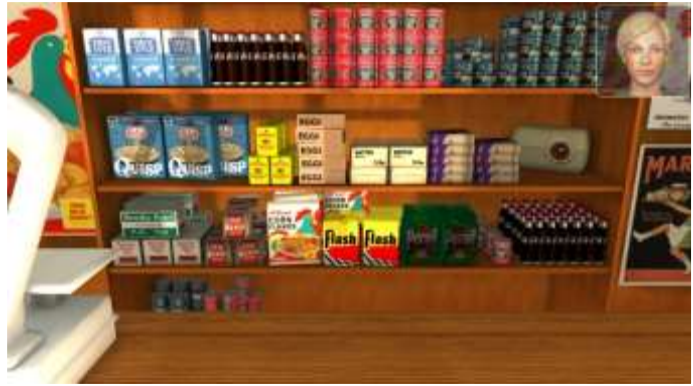

**Fig. 2.** Store scene with photorealistic Alex

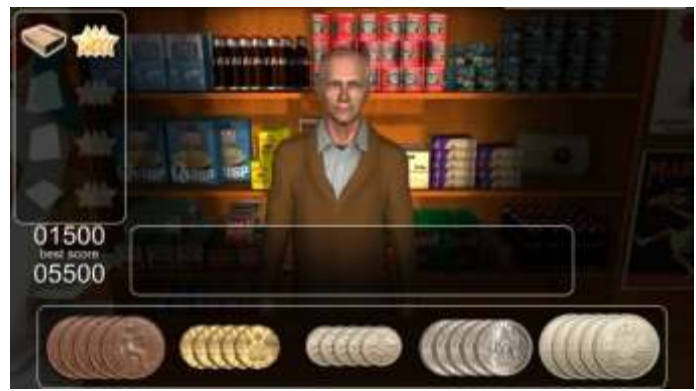

**Fig. 3.** Photorealistic shopkeeper with the inventory and coins

### 3.3   Procedure

The experiment was conducted in a dedicated laboratory setting within the university. Participants were welcomed and informed of the purpose of the study. The experiment session lasted approximately 45 minutes. After experiencing each version, participants completed a set of 7-point Likert scale usability questions as well as an exit interview to explore issues raised in the experiment. The usability questionnaire used in this evaluation is the CCIR MINERVA usability questionnaire, a standardized and validated metric for assessing usability with 18 usability statements [34].

The first scene of the game introduces the participant to Moneyworld and a tutorial introducing the old money system used throughout the game. The tutorial is introduced by a disembodied voice and the agent is not present, as the tutorial is the same for both versions and experienced once at the beginning of the session to avoid overexposure to one style over the other. After the tutorial, the game starts with a small introduction to the gameplay delivered by Alex the instructor. The gameplay is straightforward, the player is asked by Alex to buy a list of items one at a time from the shopkeeper. For each item, the player is rewarded with points and stars. The reward system is based on the evaluation of correct payment, no help required (ex. ask twice for the money), and using as fewer coins as possible in the transaction.

### 3.4 Experimental design

We used a 2 x 2 within-subjects factorial design with the following independent variables and levels:
- ECA design: Photorealistic, Animated
- Shopping list: Shopping list 1, Shopping list 2 (different shopping lists were used to avoid overexposure but no hypothesis was attached to them)

The dependent variables were the usability questionnaire and the explicit preferences. Each participant experienced both versions and was counterbalanced using a Latin square method (Table 1). In this study, a mixed methods approach was used with a combination of qualitative and quantitative research methods.

Table 1. -Latin square for 2 x 2 factorial design

| 2 x 2 Design | | ECA Design | |
|---|---|---|---|
| | | Photorealistic | Animated |
| Shopping list | 1 | P1 | A1 |
| | 2 | P2 | A2 |

The participants were allocated randomly into equal and balanced groups with all group subjects experiencing both design options.

In the photorealistic version (Fig.4, Fig.5), the user experienced and interacted with photorealistic ECAs. The photorealistic style is used to simulate reality in a way that is believable to the user. This kind of style is mainly used for real-life simulations and immersive entertainment games [35].

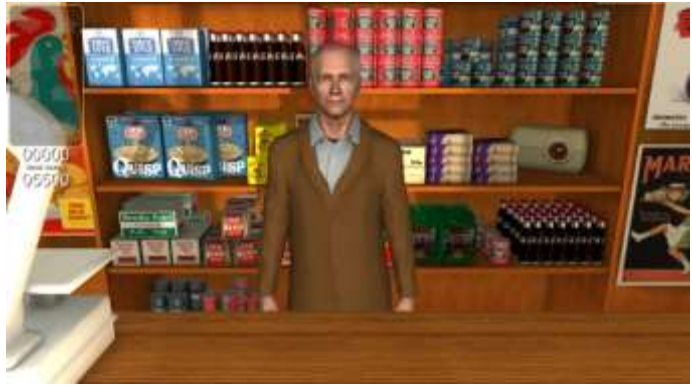

**Fig. 4.** Photorealistic shopkeeper

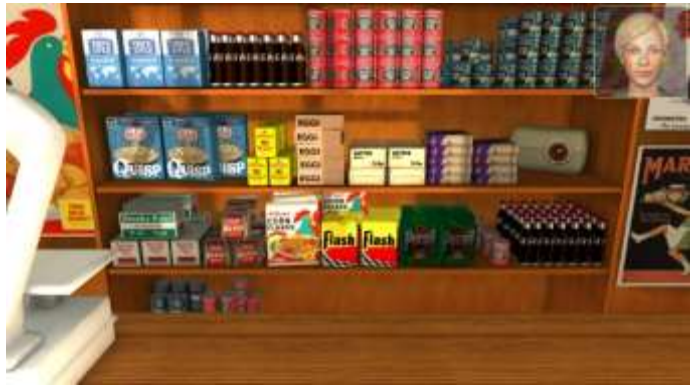

**Fig. 5.** Photorealistic Alex

In the animated version (Fig.6, Fig.7), the user experienced and interacted with animated ECAs. The animated style is based on our ability as humans to interpret abstraction quickly and effectively [36]. The animated style has been used in numerous games and gives the notion that we are not in the real world; therefore, we are not bound by its rules. The animated version was designed using the masking effect. The concept of the masking effect entails depicting characters in a stylized manner while maintaining a realistic background. Disney has successfully employed this technique for many years, achieving remarkable results in terms of character engagement [29].

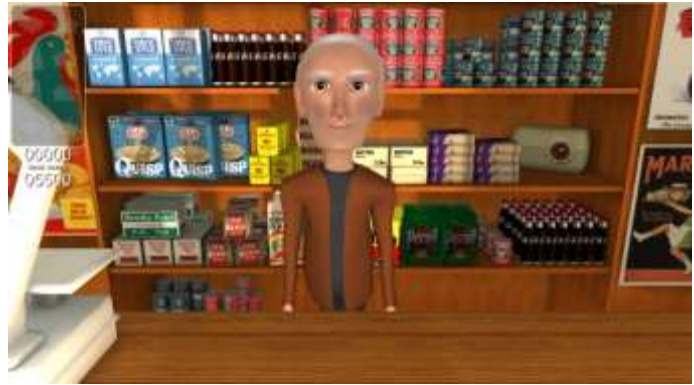

**Fig. 6.** Animated shopkeeper

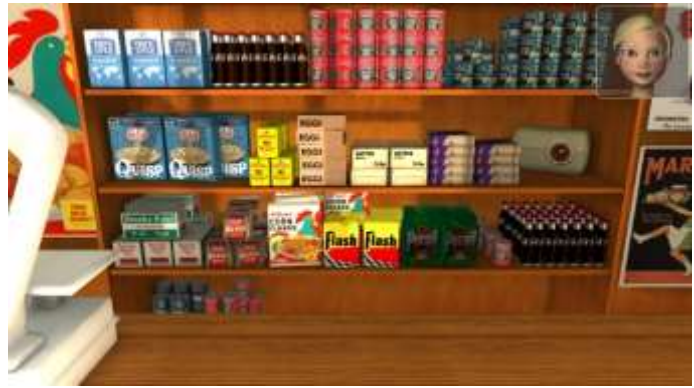

**Fig. 7.** Animated Alex

## 4  Results/findings

### 4.1  Quantitative results

**Usability Results.** An overall mean usability score was calculated from the 18 usability attribute scores for each of the two treatment groups. Overall mean scores for the questionnaire taken did not differ between the two versions. After detecting and treating the data for outliers, the overall mean for the Photorealistic version was 5.76, indicating a positive attitude toward the application. The mean score for the Animated version was similar at 5.71.

A repeated-measures ANOVA with *version* as the within-subjects factor, and *gender*, version order (*v-order*), and shopping list order (*list-order*) as between-subjects factors showed that the difference in mean attitude score was not significant. Fig.8 shows the scores for each of the 18 usability attributes for each version (PH = photorealistic, AN = animated).

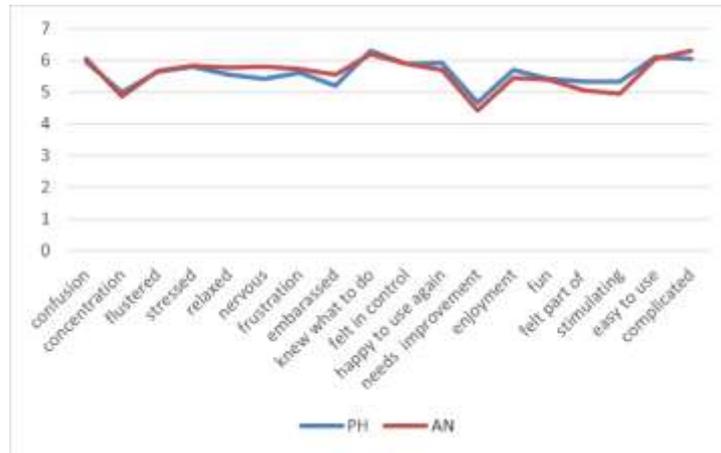

**Fig. 8.** Usability attributes for both versions

To examine any differences between the two versions for each of the individual attributes on the questionnaire, a repeated measures ANOVA was run on the mean scores; the version was the within-participants factor and the order of participation was the between-participants factor. None were found to yield significant results. There were, however, significant interactions between the versions and the order that the versions were experienced for usability statement No 2 "I had to concentrate hard to use Moneyworld." (p=0.007). This result signifies that the participant concentration levels were dependent on the order they experienced it. More specifically, this result indicates that men felt they had to concentrate significantly more when using the animated version compared to the photorealistic one while women rated the two versions similarly on this issue (Fig.9).

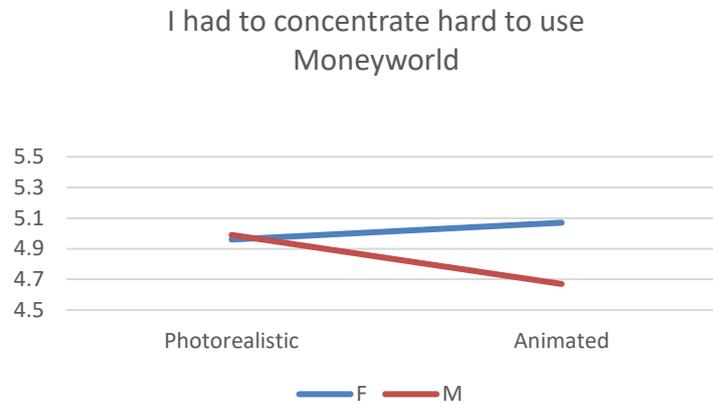

**Fig. 9.** Interaction between version and gender regarding concentration

### 4.2 Qualitative results

After experiencing each version, participants were asked to comment on their experience with the application and then specifically on each version they experienced. The exit interview consisted of both open and closed questions.

The overall perception of the tutorial was positive with 91.7% finding it helpful. It was mentioned in the comments that the tutorial was clearly explained and informative as they learned a lot about old money. Following the tutorial, 97.2% felt they understood the old money. When asked about the voice introducing the tutorial, all comments were positive. Even though this was an open question, the answers were organized and analyzed for recurring themes. In terms of what participants liked in the voice used in the tutorial, 16 responded that the voice was well-paced and clear; 10 mentioned it was friendly, welcoming, and pleasant, and the rest that the voice was good and fine.

**Shopkeeper.** During the exit interview, participants were asked "What did you think of the shopkeeper". An initial sentiment analysis was performed that was then followed by manual annotation and qualitative analysis to validate and refine the results obtained from sentiment analysis. Regarding the photorealistic shopkeeper, 19 participants had a positive impression based on their comments on the human characteristics of the ECA, saying that he was realistic, funny, polite, friendly, and kind which prompted them to be polite when talking to him. Some of their comments were:

- "Very realistic, smartly dressed. Makes you feel engaged as part of the shop."
- "Good character, naturally warm, he made me want to be polite as I would be to an elderly person."

Nine participants had neutral opinions about the shopkeeper saying he was fine or ok with comments like:

- "Eyebrows were funny, but it was OK in general."

Lastly, 8 participants made negative comments about the ECA, saying that he was creepy, strange looking, and stilted; this can be explained by the uncanny valley theory. Some of their comments were:

- "He was ok but a bit creepy looking."
- "Pretty creepy. More realistic but unappealing."

Participants were also asked their opinion about the animated shopkeeper. Based on the positive comments (7), the animated shopkeeper was funnier and more stylized which made the participants have less expectations of him with comments like:

- "Much more stylized, you know it is not real because of the exaggerated features. More successful because you know it is not real. Better facial expressions."

The neutral comments (12) were mostly about the basic and avatar-like graphics. Some of the comments were:

- "Short of standard computer graphic"
- "Basic graphic, clearly non-human"

Finally, negative comments (17) were mostly about the ECAs' proportions, cartoon-like appearance, and unpleasant disposition. A few of the comments were:

- "He looks like a puppet with a big head."
- "Quite scary, cartoony, and weird looking."

Fig.10 demonstrates the different sentiments between the two shopkeeper versions.

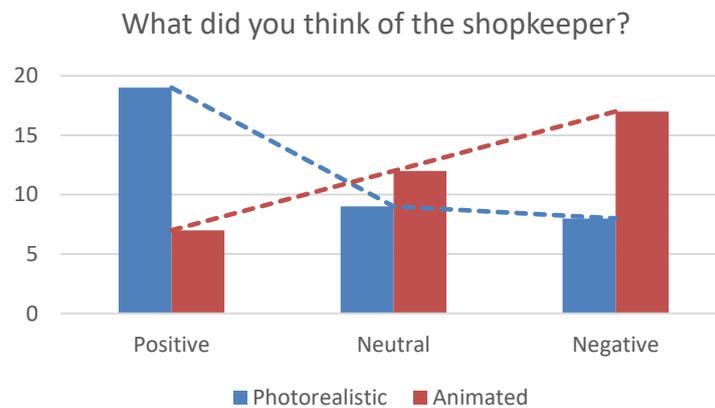

**Fig. 10.** Comparison of participant comments between the two shopkeeper versions

**Instructor.** Participants were asked about the two instructor versions as well. Regarding the photorealistic instructor, most comments were positive with 17 participants reporting that they felt she was friendlier and more realistic. Some of the positive comments were:

- "Much more human looking and friendlier."
- "Very real looking, I took information from her more seriously than the other one."

The neutral comments (13) had to do with the fact that they did not interact with her directly and did not pay too much attention to her. An example comment follows:

- "Neutral, not involved in the learning process she was the instructor.
- Didn't mind her in either version."

The negative comments (6) had to do with the lip-synching that was off and that she looked a bit robotic. A few of these comments were:

- "Felt a bit robotic, not as realistic as the shopkeeper."
- "Fine but the mouth didn't move as it was meant to."

When asked about the animated instructor, the positive comments (8) had to do with her pleasant cartoon-like appearance and the exaggerated features that made it easier to follow what she was saying. Some of the comments were:

- "Easier to watch than the first cause facial features were more eccentric."
- "It was a non-human computer-generated face but pleasant nonetheless."

The neutral comments (5) were mostly about how they did not pay as much attention to her as her role was not as interactive as the shopkeeper. One comment was:

- "I didn't notice much difference between the first and the second."

The negative comments (4) had to do with her exaggerated features that made her robotic, her face did not add to the experience or was off-putting. Some negative comments were:

- "Distracting. Doesn't look like a person."
- "Very cartoon-looking and unnatural. Maybe good for kids"

Fig.1 demonstrates the different sentiments between the two instructor versions.

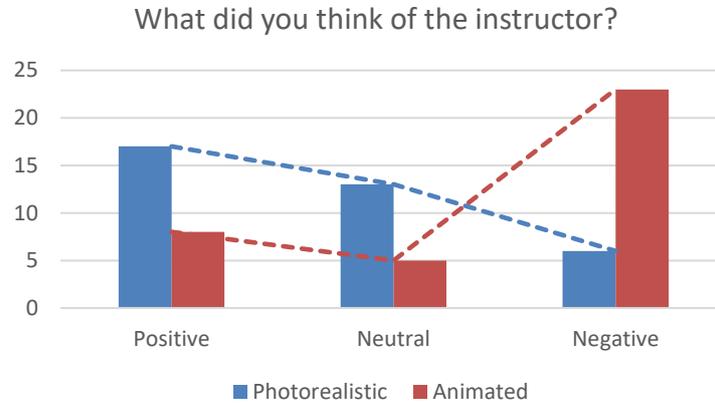

**Fig. 11.** Comparison of participant comments between the two instructor versions

### 4.3 Explicit preference

Finally, participants were asked which version of Moneyworld they preferred. Participants were asked to give their answers in terms of their first or second version experienced, and the answers were re-ordered for each version. Twenty-five out of thirty-six participants (69.4%) preferred the photorealistic version while 9 (25%) preferred the animated one and 2 (5.6%) had no preference (Fig.12).

Participants were asked to elaborate on their answers. The reasons for those who preferred the Photorealistic version were realism (10), better graphics (8), more engaging (4), more approachable and friendly (3), human-like (3), and clearer (2). Some sample comments made by participants are:

- "Felt more engaging cause the characters looked more realistic, I found it less distracting."
- "The human face makes it more approachable."
- "Felt more engaging cause the characters looked more realistic, I found it less distracting."

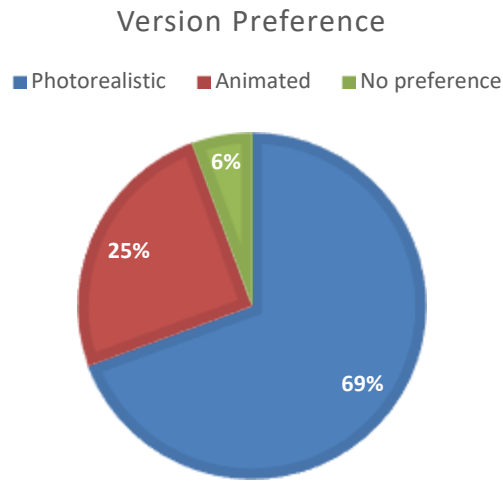

**Fig. 12.** Version preference in %

The participants who preferred the animated version justified their preference by saying that the photorealistic characters were too real and creepy (5), or they preferred the animated characters better (4). Some sample comments made by participants are:

- "Didn't feel like a test because the characters were cartoon-like."
- "Less creepy and wasn't as frustrating."
- "I preferred the look of the animated character. The first ones (photorealistic) were old, too realistic, and disconcerting. The second one fitted better with the game."

The participants were also asked which set of agents they preferred. Twenty-eight (77.8%) chose the Photorealistic agents, 7 (19.4%) the Animated agents and 1 (2.8%) had no preference (Fig.13). The Photorealistic agents were preferred due to their realism (7), better graphics (5), human-likeness (3), expressiveness (3), natural interaction, especially when using speech interaction (3), engagement (2), friendliness, personability, trust and interactivity (1). Some sample comments made by participants are:

- "More engaging, didn't feel as silly talking to realistic people."
- "Liked talking to a person rather than a cartoon."
- "See people's emotions is better and fitted better with the game."

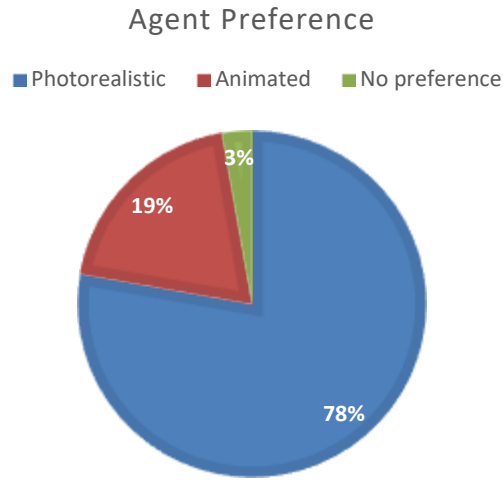

**Fig. 13.** Agent preference in %

Three people who preferred the Animated version stated that they preferred the Photorealistic characters. Their comments were:

- "You trust the realistic one more."
- "The cartoon was funnier, but I felt like I got it right with the human."
- "Bit more human. Not as creepy."

Participants justified their choice of preferring to interact with the Animated agents by stating that it was more like a game (3), more fun, and more relaxing. Some of their comments were:

- "They are characters that don't look like real people. That's the point of playing a game."
- "Polite and didn't like the photo real one in comparison. Meant to be on a computer so you expect to see computerized people more."

Only one person who preferred the Photorealistic version liked the Animated characters better and their comment was:

- "Look more fun and playful."

Regarding the agents' voices, participants were asked if they believed that the voice matched the appearance of the agent. Even though both Photorealistic and Animated agents had the same animations and voices (real actor recordings), participants stated that the voice matched the Photorealistic agents more than the Animated ones as shown in Fig.14.

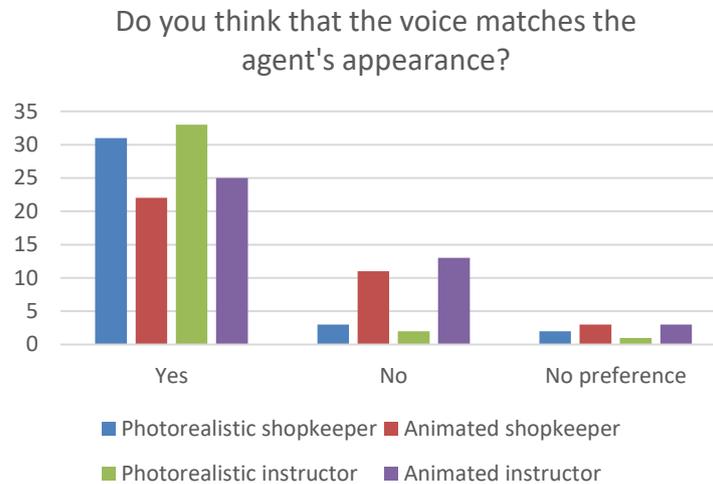

**Fig. 14.** Voice-visual perception

## 5  Discussion and limitations

The selection of agents for the photorealistic version was based on their realism. However, it is important to acknowledge the limitations of this academic project in terms of funding, technology, and expertise, which prevented the creation of hyper-realistic industrial game characters. Despite these limitations, participants still perceived the photorealistic agents as human-like and realistic, as evident from their comments. Furthermore, the limitations mentioned above also imposed a constraint on the animations, specifically the synchronization of lip movements. This aspect influenced the participants' perception of the agents and led to various effects, which will be explored and discussed in detail later in the section.

The quantitative data did not reveal a significant usability difference between the two versions with scores that suggested a relatively positive evaluation of usability for both. Regardless, there was an interesting effect between gender and version when it came to concentration levels which can be explored further in the future.

The analysis of qualitative data provided valuable insights into participants' preferences. While some participants found both the animated and photorealistic shopkeepers to be "creepy", there was a higher level of anticipation and expectations associated with the shopkeeper that had a realistic appearance. This phenomenon can be explained by the uncanny valley theory [37] which is a common risk when developing human-like interfaces. Many comments associated the animation of facial features, eyebrows, and lip-synching, with this feeling of uneasiness. According to Gratch [37], gesturing without any facial expression can look peculiar and vice versa. The same is true for the movement of hands without any involvement of the torso. Moreover, facial expressions should accomplish any attempts of emotions because, otherwise, the lack of facial involvement could detract from the expected result even if the character's speech and gestures are synched [37]. Saygin et al. [38] derived the conclusion that the effect can be based on perceptual mismatch where ultra-realistic human-like robots are expected to behave in an equally realistic human-like way. Therefore, the strength of the effect relies on how high the expectation is [4].

Overall, the photorealistic agents were perceived as friendlier, realistic, and engaging highlighting that the human-like characteristics of the ECA made that version more appealing to the participants. An interesting observation is that even though both versions had the same animations and voices, most participants stated that the voices matched the photorealistic agents better. This could be because the voices were actor recordings and not computer generated thus finding the more realistic-looking ECAs as a better fit.

Another noteworthy aspect was the effect of speech interaction with the ECAs. Several comments indicated that participants felt a greater sense of naturalness when engaging in conversation with a realistic, human-like character. This preference can be attributed to the familiarity with speech interaction in human-to-human communication. Consequently, participants tended to gravitate more towards realistic agents when replicating this form of communication as it enhances the sense of authenticity and relatability in human-computer interactions. This could be explained by the media equation theory [39, 40] and the illusion of humanness effect [3]. The media equation theory implies that people tend to interact with computers and media in an inherently social way [39]. Reeves and Nass [40] demonstrated that interactions between humans and agents exhibit comparable fundamental patterns to those observed in human-to-human interactions [29]. An extension of the media equation is the illusion of humanness. The illusion of humanness is defined as the user's involuntary reaction to a humanoid and anthropomorphic interface which gives them the perception that the system possesses human attributes and/or cognitive functions [3]. Essentially, as the agent's realism increases, so do the social and cognitive expectations placed upon it, an effect that was observed in this study as well.

## 6 Summary and conclusions

In conclusion, this study provides empirical data on the effect of design styles of ECAs within serious games. It contributes to the ECA literature by providing insights into the role of aesthetics in user interaction with serious games. It also offers design suggestions for ECAs within serious games by shedding light on the preferences and experiences of users when interacting with photorealistic and animated ECAs. A 2 x 2 within-subjects factorial design was utilized with participants experiencing both conditions. While the quantitative results indicated that there were no significant differences in the overall usability scores, qualitative findings offered valuable insights into the way participants perceived the ECAs. Comments on the photorealistic shopkeepers were mostly positive due to their realism, human likeness, and engagement despite the un-canny feelings evoked due to the animation. On the other hand, the animated shopkeeper was praised for its funny and stylized appearance but also received negative comments on its disproportionate features and cartoon-like appearance. Regarding the instructors, the photorealistic version was perceived as more friendly and realistic, while the animated version received mixed comments on its cartoon-like appearance and exaggerated features.

Overall, the results of the comparative evaluation of animated versus photorealistic ECAs in a serious game generate insights into best practices for designing ECAs in serious game applications. These results indicate that the visual appearance of ECAs can impact user preferences, with photorealistic agents generating more favorable reactions among participants. However, these preferences may not lead to significant variations in the overall usability of the agents. Nonetheless, while the results indicated a preference for photorealistic ECAs in serious games, ECAs should be evaluated in various contexts to establish this effect. This research aims to lay the groundwork for future investigations into the impacts of ECA realism in serious games across diverse contexts.

**Acknowledgments.** I would like to thank Dr. Nancy Gunson, Dr. Hazel Morton, Professor Mervyn Jack, and Dr. Simon Doolin for their invaluable input and assistance. This research was funded by Lloyds TSB.